\documentstyle[aps,12pt]{revtex}
\input psfig
\newcommand{\beq}[2]{\begin{equation}#1\label{#2}\end{equation}}
\newcommand{\ceq}[1]{(\ref{#1})}
\newcommand{\mbd}[1]{\mbox{\bf #1}}
\newcommand{\bp}{\mbd{p}}

\newcommand{\br}{\mbd{r}}

\newcommand{\bA}{\mbd{A}}
\newcommand{\bD}{\mbd{D}}
\newcommand{\bB}{\mbd{B}}
\newcommand{\bC}{\mbd{C}}

\newcommand{\tgm}{G_{\{m\}}(\{\br\},\{L\};\{\br'\},0)}

\newcommand{\tgl}{G_{\{\lambda\}}(\{\br\},\{L\};\{\br'\},0)}
\newcommand{\tgz}{G_{\{\lambda\}}(\{\br\},\{\br'\},\{ z\})}
\newcommand{\ttgz}{G_\lambda(\{\br\},\{\br'\},\{ z\})}
\newcommand{\tgmz}{G_m(\{\br\},\{\br'\},\{ z\})}

\newcommand{\tglfc}{G(\br_i,L_i;\br_i',0|\phi_i,\bC^{(i)})}
\newcommand{\tgzfc}{G(\br_i,\br_i';z_i|\phi_i,\bC^{(i)})}

\newcommand{\Aij}{\bA^{(i)}_{(j)}}
\newcommand{\Bij}{\bB^{(i)}_{(j)}}
\newfont{\mbld}{cmbsy10 scaled 800}
\begin{document}
\title{TOWARD\ A\ FIELD\ THEORETICAL\ DESCRIPTION\ OF\ TOPOLOGICALLY\
LINKED\ POLYMERS}
\author{Franco Ferrari$^{(1)}$ and Ignazio Lazzizzera$^{(1)(2)}$\\
{$^{(1)}$\it Dipartimento di Fisica, Universit\'a di Trento, 38050 Povo (TN),
Italy.}\\
{$^{(2)}$\it INFN, Gruppo Collegato di Trento, Italy.}}
\date{March 99}
\maketitle
\vspace{-4.5in} \hfill{Preprint UTF XXX/YY} \vspace{4.4in}
\begin{abstract}
In this work a field theoretical model is constructed
to describe the statistical
mechanics of an arbitrary number of topologically linked polymers
in the context of the analytical approach of Edrwards.
As an application, the effects of the topological interactions are studied in
the one loop approximation. A natural way to include in the treatment also
more sophisticated link invariants than the Gauss linking number
is outlined in the Conclusions.
\end{abstract}
\section{Introduction}
Long flexible polymer molecules provide an ideal environment
to study the effects of topology in statistical mechanics and condensed matter
physics, but so far
a comprehensive theory of
topologically linked polymers does not exist. Until now, only the physics of
unentangled open chains has been understood to a satisfactory
extent \cite{degennes}.
In this letter the problem is tackled in the
context of the
so-called analytical approach, consisting in a
number of different methods based on Edwards pioneering work
\cite{edwards}, in
which the fluctuations of each polymers are treated as a random walk
and described via a path integral in the Wiener sense.
Analytical approaches of this kind have already
been applied
to several complex situations
\cite{otvi}-\cite{brereton} and 
seem very promising in
explaining the
physics of topological
entangled polymers.
Despite of that, they are affected by two main limitations.
First of all, the Gauss linking number, a  weak
topological invariant, is used to
distinguish the topological states of the system.
Unfortunately, 
more sophisticated knot invariants
like those described in \cite{kauffmann} and
\cite{jones} cannot be applied in practice, since they
have no
immediate relation to the physical conformation of the polymer \cite{otvi}.
On the other side, one of the advantages of the analytical approach is
to provide a field theoretical description of a system of polymers,
establishing in this way a connection to the theory of critical phenomena in
condensed matter and high energy physics. However, a mapping
from the statistical mechanical problem of polymers to field theories
has been realized only in a limited number of cases, mainly
concerning the fluctuations of a single test polymer in a background of
static polymers or fixed obstacles. Remarkable exceptions occur
when two polymers are very close \cite{tanakaii},
\cite{brereton},
a situation which is particularly relevant in the study
of DNA molecules \cite{moka}, or when the number of polymers becomes
large \cite{vilbreav}.
Despite of these successes, the exact analytical treatment of a system
of $N$ polymers has been a formidable problem already mentioned in the
pioneering work on topological entanglement of Brereton and Shah \cite{brsh}
and remained unsolved until the present days.

To solve it, we construct a Chern-Simons (C-S) 
\cite{chernsimons}
based model, which
generalizes a previous work of the authors
in which the two polymer case was discussed
\cite{felaone}.
With respect to ref. \cite{felaone}, the main difficulty is that
one needs $\frac{N(N-1)}2$ independent topological numbers
to distinguish the possible topological
states of $N$ polymers according to the Gauss link invariant.
As a consequence, one has to find a suitable C-S field
theory
which decouples the actions of the
polymers as in \cite{felaone} and simultaneously is able to accommodate the
huge amount of topological configurations allowed by the system.

The material contained in this paper will be divided as follows.
In the next Section a topological Ginzburg-Landau model is
constructed which describes the fluctuations of a system of $N$ topologically
linked polymers.
The topological relations are specified using the
Gauss linking number. Section III contains a perturbative study of the
case $N=2$. In particular, the one loop effective potential is computed
proving that the topological interactions do not influence the
critical behavior of the system in this approximation.
Moreover, it is shown that the second topological moment can be
exactly computed in terms of one loop Feynman diagrams. Finally, a
detailed discussion of the results has been given in Section IV.
In that Section a natural method to include higher order topological
invariants is outlined. To this purpose, a new C-S field
theory is defined, which is abelian but exhibits cubic interactions
among the C-S fields. In this way such theory can generate
radiative corrections containing knot invariants with the same
mechanism acting in the case of nonabelian  string theories \cite{witten},
\cite{agua}.
\section{The $N-$polymers problem}
Let $P_1,\ldots,P_N$ be a set of topologically linked polymers,
describing in the space the trajectories $\Gamma_i$, $\enskip
i=1,\ldots,N$ defined as follows:
\beq{\Gamma_i=\left\{\br_i(s_i)\left | 0\le s_i\le
L_i\right. ; 
\br_i(0)=\br'_i, \br_i(L_i)=\br_i \right\}} {trajectdef}
where $L_i$ denotes the contour length of $\Gamma_i$. The polymers can
be either open or closed ($\br_i'=\br_i$).
The topological states will be distinguished using the Gauss linking
number:
\beq{
 \chi(\Gamma_i,\Gamma_j)\equiv 
 \frac{1}
 {4\pi}
 \int_0^{L_i}
 \int_0^{L_j}
 d\br_i(s_i)\times
 d\br_j(s_j)\cdot
    \frac
            {
              \left(
                     \br_i(s_i)-
                     \br_j(s_j)
              \right)
             }
             {
              |\br_i(s_i) - \br_j(s_j)|^3
             }
}
{glinv}
where
$\chi(\Gamma_i,\Gamma_j)$ takes integer values $m_{ij}$
if $\Gamma_i,\Gamma_j$, $i\ne j=1,\ldots,N$, are closed
trajectories. 
To describe the statistical mechanics of the polymers we define the
configuration probability $\tgm$.
This function measures the probability that the trajectories
$\Gamma_i$ fulfill the constraints:
\beq{
 \chi(\Gamma_i,\Gamma_j) = m_{ij}\qquad\qquad\qquad i\ne j = 1,\ldots,N
}{constraint}
and have extrema (in the open case) at the points $\br_i'$ and $\br_i$,
$i=1,\ldots,N$.
In our notations $\{m\}$ denotes the $n\times n$ symmetric matrix of
topological numbers with elements
\beq{
m_{ij}=m_{ji}\qquad \mbox{\rm for}\qquad i\ne j\qquad\qquad ;\qquad\qquad
m_{ii}=0
}{mijdef}
while $\{\br\}= \br_1,\ldots,\br_N$, $\{L\}=L_1,\ldots,L_N$ etc.
Starting from the path integral approach of Edwards \cite{edwards},
one obtains the
following expression of
$\tgm$:
\beq{
\tgm =\int_{-\infty}^{+\infty}\left[\prod_{i=1}^{N-1}\prod_{j=2\atop
j>i}^N
\frac{d\lambda_{ij}}{2\pi}e^{-i\lambda_{ij}m_{ij}}\right]
\tgl}{ftgmgl}
where
\beq{\tgl = \int_{\br_1'}^{\br_1}{\cal D} \br_1(s_1) \ldots
\int_{\br_N'}^{\br_N}{\cal D} \br_N(s_N)
\mbox{\rm exp}\left\{-\left( {\cal A}_0+{\cal A}_{ev}+{\cal A}_{top}\right)
\right\}}{glambda}
In the above equation the action
\beq{
{\cal A}_0=\frac 3{2a}\sum_{i=1}^N\int_0^{L_i}\dot{\br}^2_i(s_i)}{frw}
is that of a particle free random walk corresponding to a trajectory
$\Gamma_i$, while
\beq{{\cal A}_{ev}=\frac 1{2a^2}\sum_{i,j=1}^N
\int_0^{L_i}
ds_i\int_0^{L_j}
ds'_jv^0_{ij}\delta^{(3)}(\br_i(s_i)-\br_j(s'_j))}{exvolact}
takes into account the excluded volume interactions.
For convenience, $v^0_{ij}$ has been defined as follows:
\beq{
v^0_{ij}=
\left\{
\begin{array}{c c}
\tilde v^0_{ij}&\mbox{\rm for}\enskip i=j\cr
\tilde v^0_{ij}/2&\mbox{\rm for}\enskip i\ne j\cr
\end{array}
\right.\qquad\qquad \tilde v^0_{ij}=\tilde v^0_{ji}}{vzij}
where the $\tilde v^0_{ij}$ are coupling constants with the dimension of a
volume.
Finally, the topological constraints \ceq{constraint} are responsible of the
topological term:
\beq{
{\cal A}_{top}=i\sum_{i=1}^{N-1}\sum_{j=2\atop
j>i}^N\chi(\Gamma_i,\Gamma_j) \lambda_{ij}}{topintterm}
coming from the well known Fourier transform of the Dirac
$\delta-$function:
\beq{
\delta(m_{ij}-\chi(\Gamma_i,\Gamma_j))=\int_{-\infty}^{+\infty} 
\frac{d\lambda_{ij}}{2\pi}\mbox{\rm exp}\left\{-i\left[m_{ij}-
\chi(\Gamma_i,\Gamma_j)\right]\lambda_{ij}\right\}}{ddffrep}
To express the configurational probability \ceq{glambda} in terms of
fields, we introduce $N$ gaussian scalar fields $\phi_i(\br)$,
$i=1,\ldots,N$ 
and $N(N-1)$ free C-S fields
$\Aij$ and $\Bij$, with $i=1,\ldots,N-1$, $j=2,\ldots,N$ and
$j>i$.
The action of the scalar fields is given by:
\beq{{\cal A}_{\phi}=\frac{a^2}2\sum_{i,j=1}^N\int
d^3\br\phi_i[(v^0)^{-1}]^{ij}\phi_j}{aphi}
while C-S one is given by:
\beq{
S_{CS} =\frac \kappa{4\pi} \int d^3\br\sum_{i=1}^{N-1}
\sum_{j=2\atop j>i}^N \Aij\cdot(\nabla\times\Bij)}{csaction}
The C-S theory will be quantized in the Landau gauge, where the fields
$\Aij$ and $\Bij$ are completely transverse.
It turns out that the following relation is valid for the excluded volume term:
\beq{e^{-{\cal A}_{ev}}=\int{\cal D}\phi_1\ldots{\cal D}\phi_N\enskip
\mbox{\rm exp}\left\{-{\cal A}_{\{\phi\}} -i \sum_{i=1}^N\int d^3\br
J_i(\br)\phi_i(\br)\right\}}{evitofa}
where
\beq{
J_i(\br)=\int_0^{L_i}ds_i\delta^{(3)}(\br-\br_i(s_i))}{currphi}
To treat the topological interaction it will be convenient to
define the following linear combinations of the fields
$\Aij$ and $\Bij$:
\beq{
\bC^{(1)}=\sum\limits_{j=2}^N\alpha^{(1)}_{(j)}\bA^{(1)}_{(j)}}{cone}
\beq{
\bC^{(i)}=\sum\limits_{j=3\atop j>i}^N\alpha^{(i)}_{(j)}\bA^{(i)}_{(j)}
+\sum\limits_{j=1\atop j<i}^{N-2}\beta^{(j)}_{(i)}\bB^{(j)}_{(i)}
\qquad\qquad i=2,\ldots,N-1}{ctnmo}
and
\beq{
\bC^{(N)}=\sum\limits_{i=1}^{N-1}\beta_{(N)}^{(i)}\bB^{(i)}_{(N)}}{cenne}
The parameters $\alpha^{(i)}_{(j)}$ and $\beta^{(i)}_{(j)}$ depend
on the C-S coupling constant $\kappa$ and on the matrix elements
$\lambda_{ij}$ as follows:
\beq{\alpha^{(i)}_{(j)}=\frac \kappa{4\pi}\qquad\qquad\qquad
\beta^{(i)}_{(j)}=\lambda_{ij}}{coefficienti}
In terms of these fields $\bC$ it is now possible to state the
Gaussian identity:
\beq{
\int{\cal D\bA}{\cal D\bB}\enskip\mbox{\rm exp}\left\{-i S_{CS}-i\sum_{i=1}^N
\int_0^{L_i}\bC^{(i)}(\br(s_i))d\br(s_i)\right\}=
\mbox{\rm exp}\left\{-i\sum_{i=1}^{N-1}\sum_{j=2\atop j>i}^N\lambda_{ij}
\chi(\Gamma_i,\Gamma_j)\right\}}{relfond}
Eq. \ceq{relfond}, where we have used the compact notation:
\beq{{\int\cal D\bA}{\cal D\bB}\equiv\int\prod_{i<j=1}^N
{\cal D}\Aij{\cal D}\Bij}
{posone}
relates the topological term \ceq{topintterm} to the amplitude of $N$
holonomies and will be central in the subsequent discussion.
Substituting eqs. \ceq{relfond} and \ceq{evitofa}
in \ceq{glambda}, the configurational probability
$\tgl$ becomes:
\beq{
\tgl=\langle\prod_{i=1}^N\tglfc\rangle_{\{\phi\},\{\bA\},\{\bB\}}}
{factorized}
In the above formula the functions $\tglfc$ are given by:
\beq{
\tglfc=\int_{\br'_i}^{\br_i}
{\cal D}\br_i(s_i)
\mbox{\rm exp}
\left\{
-\int_0^{L_i}ds_i{\cal L}_{\phi_i}
(\br_i(s_i))
-i\int_0^{L_i}ds_i\dot{\br}_i(s_i)
\cdot\bC^{(i)}(\br_i(s_i))
\right\}
}
{gsimple}
where
\beq{
{\cal L}_{\phi_i}
(\br_i(s_i))= 
\frac 3{2a}
\dot{\br}_i^2(s_i)+i\phi_i(\br_i)}
{lphi}
Moreover, everything inside the bracket
$\langle\qquad\rangle_{\{\phi\},\{\bA\},\{\bB\}}$
must be averaged with respect to the fields
$\{\phi\},\{\bA\}$ and $\{\bB\}$ by means of eqs. \ceq{evitofa} and
\ceq{relfond}.
As we see from eqs. \ceq{factorized}-\ceq{lphi} the polymer trajectories
$\Gamma_1,\ldots,\Gamma_N$  are completely decoupled before averaging
over the auxiliary fields. Furthermore, each of the factors
$\tglfc$ appearing inside the bracket
$\langle\qquad\rangle_{\{\phi\},\{\bA\},\{\bB\}}$
coincides formally with the evolution
kernel of a particle subjected to random walk (in the Wiener sense of
the path-integral formulation of statistical mechanics) and immersed in the
electromagnetic field $(i\phi_i,\bC^{(i)})$.
Thus $\tglfc$ satisfies the pseudo-Schr\"odinger equation \cite{tanaka}:
\begin{equation}
\left[\frac\partial{\partial{L_i}}
-\frac a 6\mbd{D}_i^2+i\phi_i\right]
G(\br_i, L_i;\br'_i,0|\phi_i,\bC^{(i)})=
\delta(L_i)\delta(\br_i-\br'_i)
\label{pseschroe}
\end{equation}
The covariant derivatives $\bD_i$ are given by:
\beq{\bD_i=\nabla+i\bC^{(i)}\qquad\qquad\qquad i=1,\ldots,N}{covdevdef}
At this point it is convenient to consider the Laplace transform of the
configurational probability $\tgl$ with respect to the polymer lengths
\cite{tanaka}:
\beq{
\tgz=\int_0^{+\infty}d_{L_1}\ldots\int_0^{+\infty}d_{L_N}
\mbox{\rm exp}\left\{-\sum_{i=1}^Nz_iL_i\right\}\tgl}{gllapltransf}
Each variable $z_i$ plays the role of a Boltzmann-like factor
governing the distribution length of the $i-$th polymer.
Applying the above Laplace transformations to the expression of
$\tgl$  given by eq. \ceq{factorized}, one finds:
\beq{\tgz=\langle\prod_{i=1}^N\tgzfc\rangle_{\{\phi\},\{\bA\},\{\bB\}}}
{facttwo}
where the functions $\tgzfc$ 
are the Laplace transforms of $\tglfc$. 
Therefore, they obey the {\it stationary} pseudo-Schroedinger
equations:
\beq{\left[z_i-\frac a 6\bD_i^2+i\phi_i\right]
\tgzfc=\delta(\br_i-\br'_i)}{statpseschroe}
The solution of eq. \ceq{statpseschroe} can be given in terms of
second quantized fields fields $\psi_i^*,\psi_i$, $i=1,\ldots,N$:
\begin{equation}
\tgzfc=
{\cal Z}_i^{-1}\int
{\cal D}\psi_i
{\cal D}\psi^*_i
\psi_i(\br_i)
\psi^*_i(\br'_i)
e^{-F[\psi_i]}
\label{gsimplesq}
\end{equation}
where
$F[\psi_i]$ represents the Ginzburg-Landau free
energy of a superconductor in a fluctuating magnetic field:
\begin{equation}
F[\psi_i]=\int d^3\br\left[
\frac a 6|{\mbd D}_i\psi_i|^2 +
(z_i+i\phi_i)|\psi_i|^2
\right]
\label{freenergy}
\end{equation}
and ${\cal Z}_i$ is the partition function of the system:
\begin{equation}
{\cal Z}_i=\int
{\cal D}\psi_i
{\cal D}\psi^*_i
e^{-F[\psi_i]}
\label{ztpartfun}
\end{equation}
In order to eliminate the auxiliary fields $\phi_i$, we exploit the
method of replica \cite{repltrick}.
Let $\psi_i^{*\omega_i},\psi_i^{\omega_i}$ be a set of
replica fields with $\omega_i=1,\ldots,n_i$ and $i =1,\ldots,N$, that
form the $n_i-$ples
$\Psi_i=(\psi_i^1,\ldots,\psi_i^{n_i})$ and
$\Psi_i^*=(\psi_i^{*1},\ldots,\psi_i^{*n_i})$.
In terms of these fields the Green functions $\tgzfc$ become:
\begin{equation}
\tgzfc=
\lim_{n_i\to 0}\int
{\cal D} \Psi_i
{\cal D} \Psi_i^*
\psi_i^{\overline{\omega}_i}(\br_{i})
\psi_i^{*\overline{\omega}_i}
(\br'_i)\enskip
e^{-\sum\limits_{\omega_i=1}^{n_i}
F[\psi_i^{\omega_i}]}
\label{replgsimp}
\end{equation}
where the $\overline{\omega}_i$'s are arbitrary integers chosen in the range
$1\le \overline{\omega}_i\le n_i$.
According to the replica method, we also assume that the limit for $n_i$
going to zero commutes with the functional integrations over the
C-S fields and the auxiliary fields $\phi_i$.
Substituting eq. \ceq{replgsimp} in \ceq{facttwo} and performing the Gaussian
integration over the $\phi_i$, one obtains:
\[
\tgz=
\]
\begin{equation}
\lim_{n_1,\ldots,{n_N\to 0}}
\int{\cal D}\bA{\cal D}\bB
\prod_{i=1}^N\left[
{\cal D} \Psi_i
{\cal D} \Psi_i^*
\right]
\prod_{j=1}^N\left[\psi_j^{\overline{\omega}_j}(\br_j)
\psi_i^{*\overline{\omega}_i}(\br'_i)
\right]\enskip\mbox{\rm exp}
\left\{-{\cal A}
(
\{\Psi\},\{\bA\},\{\bB\})
\right\}
\label{prefin}
\end{equation}
where the polymer free energy
${\cal A}
(
\{\Psi\},\{\bA\},\{\bB\})$ is given by:
\[
{\cal A}(\{\Psi\},\{\bA\},\{\bB\})=iS_{CS}+
\]
\begin{equation}
\sum_{i=1}^N
\int d^3\br\left[\frac
a6|\mbd{D}_i\Psi_i|^2+z_i|\Psi_i|^2\right]
+\sum_{i,j=1}^N
\frac{v^0_{ij}}{2a^2}
|\int d^3 \br|\Psi_i|^2
|\Psi_j|^2
\label{preaction}
\end{equation}
From the above equation we see that the fields 
$\Psi_i,\Psi_i^*$ have the dimension of a mass, which is not the canonical
dimensionality of the scalar fields in three dimensions.
To remedy, we introduce a new mass parameter $M$ \cite{kleinert}.
It is nice to see that in this way it is possible to define
an analog of the Planck constant:
\beq{
\hbar=\frac {Ma}3}{htagliata}
Performing the field rescaling
\beq{\Psi_i'=\sqrt{\frac M2}\Psi_i\qquad\qquad\qquad\Psi_i^{\prime*}
=\sqrt{\frac M2}\Psi_i^*}{rescaling}
and working in units in which $\hbar=1$, the action \ceq{preaction}
becomes:
\[{\cal A}_N(\{\Psi'\},\{\bA\},\{\bB\})=iS_{CS}
+\sum_{i=1}^N\int d^3\br \left[\Psi_i^{\prime*}
(-\bD^2_i+m_i^2)\Psi_i'\right]+
\]
\beq{
\sum_{i,j=1}^N \frac {2M^2v^0_{ij}}{a^2}
\int d^3\br |\Psi_i'|^2|\Psi_j'|^2}{finalaction}
where the covariant derivatives are defined as before by eq. \ceq{covdevdef}
and
\beq{m_i^2=2Mz_i}{massdefs}
From now on, we will always use the rescaled action
\ceq{finalaction} and the prime indices carried by the complex scalar fields
will be dropped.
\section{The two polymers case}
In the two polymers case the matrix $m_{ij}$ of topological numbers
is replaced by a single topological number $m$ which describes the
intersections of the trajectories $\Gamma_1$ and $\Gamma_2$.
Its Fourier conjugate variable will be called $\lambda$.
It will also be convenient to put:
\beq{
\bA^{(1)}_{(2)}=\bA^{(1)}\qquad\qquad\qquad
\bB^{(1)}_{(2)}=\bA^{(2)}}
{twofields}
and
\beq{
\alpha^{(1)}_{(2)}=\gamma_1\qquad\qquad\qquad
\beta^{(1)}_{(2)}=\gamma_2}{parachoice}
In this new notation the action \ceq{finalaction} becomes:
\[{\cal A}_2(\{\Psi\},\{\bA\})=i\frac{\kappa}{4\pi}\int d^3\br
\bA^{(1)}\cdot(\nabla\times \bA^{(2)})
+\sum_{i=1}^2\int d^3\br \left[\Psi_i^*
(-\bD^2_i+m_i^2)\Psi_i\right]
\]
\beq{
+\sum_{i,j=1}^2 \frac {2M^2v^0_{ij}}{a^2}
\int d^3\br |\Psi_i|^2|\Psi_j|^2}{twoaction}
where
\beq{
\bD_i=\nabla+i\gamma_i\bA^{(i)}}
{twocovdevs}
Accordingly, the configurational probability $\tgz$ can be written
as follows in the two polymers case:
\[ \ttgz=
\]
\begin{equation}
\lim_{n_1,{n_2\to 0}}
\int
\prod_{i=1}^2\left[{\cal D}\bA^{(i)}
{\cal D} \Psi_i
{\cal D} \Psi_i^*
\right]
\prod_{j=1}^2\left[\psi_j^{\overline{\omega}_j}(\br_j)
\psi_i^{*\overline{\omega}_i}(\br'_i)
\right]\enskip\mbox{\rm exp}
\left\{-{\cal A}_2
(
\{\Psi\},\{\bA\})
\right\}
\label{tprefin}
\end{equation}
Since we are interested in the effects of the topological interactions,
we neglect the excluded volume interactions
setting $v_{ij}^0=0$ in the remaining of this Section.
\begin{figure}
\vspace{1.5truein}
\includegraphics{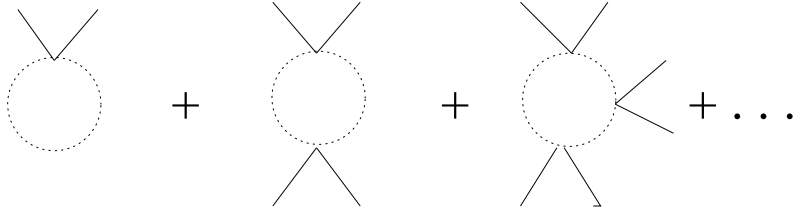}
\vspace{0.25in}
\caption{Non-vanishing contribution to the one-loop effective potential
due to the topological interactions. Solid lines represent the charged
scalar bosons and dashed lines Chern-Simons vector fields.}
\label{topintdia}
\end{figure}
Tho study the above topological Ginzburg-Landau model at one loop order
we exploit the method of effective potential \cite{cowe}.
To this purpose one has to compute only the diagrams of fig. \ceq{topintdia}.
\begin{figure}
\vspace{1.0truein}
\includegraphics{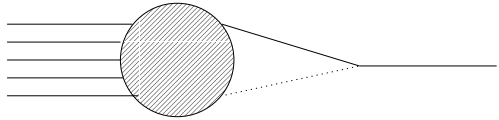}
\vspace{0.25in}
\caption{Diagrams containing at least one
vertex of the coupling  $\bA^\tau\Psi_\tau\Psi^*_\tau$.
These contributions to the effective potential
are identically zero in the Landau gauge.}
\label{vandiags}
\end{figure}
The other possible diagrams are in fact of the form given in fig.
\ceq{vandiags} and are suppressed in the Landau gauge \cite{cowe}.
First of all
one needs to derive the Feynman rules starting from the action
\ceq{twoaction}. Here we just state the result:
\beq{\langle A^{(2)}_\mu(\bp)A^{(1)}_\nu(-\bp)\rangle =\frac {4\pi}{\kappa}
\epsilon_{\mu\nu\rho}\frac {p^\rho}{\bp^2}\qquad\qquad\qquad
\mu,\nu,\rho=1,2,3}{abprop}
\beq{\langle
\psi^{\omega_i}_i(\bp)\psi_j^{*\omega_j}(-\bp)\rangle
=\frac{\delta^{\omega_i\omega_j}\delta_{ij}}{\bp^2}
\qquad\qquad\qquad i,j=1,2}{psipsibprop}
\beq{\langle\psi_i^{\omega_i}(\bp_1)\psi_i^{*\omega_i}(-\bp_2)A_\mu(\bp_3)
\rangle=
2\gamma_i\delta^{(3)}(\bp_1-\bp_2+\bp_3)(p_2)_\mu}{oaipsipsiprop}
\beq{\langle\psi_i^{\omega_i}(\bp_1)\psi_i^{*\omega_i}(\bp_2)A^{(i)}_\mu(\bp_3)
A^{(i)}_\nu(\bp_4)\rangle=
\gamma_i^2\delta_{\mu\nu}
\delta^{(3)}(\bp_1+\bp_2+\bp_3+\bp_4)}{aiajpsipsiprop}
After some calculations,
one obtains the following expression of the effective
potential at one loop order:
\beq{
{\cal V}_{eff}(\{\Psi\})=\sum_{i=1}^2m_i^2|\Psi_i|^2
-\frac{2\lambda^2\Lambda}{\pi^2}|\Psi_1|^2|\Psi_2|^2
+\frac{4\pi}3|\lambda|^3(|\Psi_1||\Psi_2|)^3}{effpotol}
In eq. \ceq{effpotol}
$\Lambda$ is an ultraviolet cut-off with the dimension of a mass.
Due to the fact that the microscopic scale of our model is given by the step
length $a$, phenomena whose spatial extensions are smaller than $a$
have no meaning \cite{tanaka}, so that we can put:
\beq{\Lambda\sim a^{-1}}{lambdaset}
From eq. \ceq{effpotol} one sees that the total effect of the topological
forces at one loop is to influence the strength of the repulsions among the
polymers. In particular, topological forces do not affect the scaling
behavior of the system.
Indeed, only the last term appearing in the second member of
eq. \ceq{effpotol} may influence 
the critical behavior of the topological
Ginzburg-Landau model \ceq{finalaction}, but it
is of higher order in
the numbers of replicas $n_1$ and $n_2$. Therefore, it does not contribute in
the configurational probability \ceq{prefin}, after the limit $n_1,n_2\to 0$
is taken.

The simplest observable that can be computed within the above approach
is the second topological moment
\beq{\langle m^2\rangle=
\frac{\int_{-\infty}^\infty
m^2\tgmz}{\int_{-\infty}^\infty
\tgmz}}{sectopmom}
Here $\tgmz$ is easily evaluated by taking the inverse Fourier
transformation with respect to $\lambda$ of the configurational probability
\ceq{tprefin}. This has been done in ref. \cite{felaone}.
In the case of the second topological moment one finds
after some calculations:
\[
\langle m^2\rangle=
\lim_{n_1,{n_2\to 0}}{\cal Z}_0^{-1}\int
\prod_{i=1}^2\left[{\cal D}\bA^{(i)}
{\cal D} \Psi_i
{\cal D} \Psi_i^*
\right]
\prod_{j=1}^2\left[\psi_j^{\overline{\omega}_j}(\br_j)
\psi_i^{*\overline{\omega}_i}(\br'_i)
\right]\enskip\mbox{\rm exp}
\left\{-{\cal A}_0\right\}
\]
\beq{\times
\left[\left(\int d^3\br \Psi_2^*\bA^{(2)}\cdot\nabla\Psi_2\right)^2
+\frac 12\int d^3\br|\Psi_2|^2\bA^{(2)}\cdot\bA^{(2)}\right]}
{mtfinal}
where
\beq{
{\cal A}_0
=i\frac{\kappa}{4\pi}\int d^3\br
\bA^{(1)}\cdot(\nabla\times \bA^{(2)})
-\int d^3\br\Psi_1^2(\bD_1^2-m_1^2)\Psi_1-
\int d^3\br \Psi_2^*(\Delta-m_2^2)\Psi_2}
{zerofenergy}
and
\beq{
{\cal Z}_0=
\int
\prod_{i=1}^2\left[{\cal D}\bA^{(i)}
{\cal D} \Psi_i
{\cal D} \Psi_i^*
\right]
\prod_{j=1}^2\left[\psi_j^{\overline{\omega}_j}(\br_j)
\psi_i^{*\overline{\omega}_i}(\br'_i)
\right]\enskip\mbox{\rm exp}
\left\{-{\cal A}_0\right\}}{zeropafu}
Let us notice that in the action ${\cal A}_0$
the field $\bA^{(2)}$ plays the role
of a Lagrange
multiplier. As a consequence, after
performing the integration in $\bA^{(2)}$ in eq. \ceq{zeropafu},
the field $\bA^{(1)}$ becomes trivial,
so that the path integral ${\cal Z}_0$ actually
describes the fluctuations of two free polymers.
\begin{figure}
\vspace{1.4truein}
\includegraphics{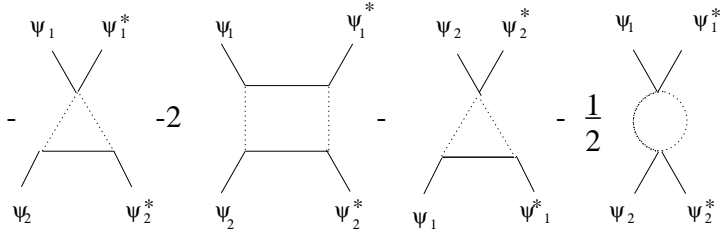}
\vspace{0.25in}
\caption{This figure shows all the diagrams contributing to the
second topological moment together with the weights they appear
in the expression of $\langle m^2\rangle$.}
\label{momentcont}
\end{figure}
The path integral appearing in the numerator of eq. \ceq{mtfinal}
is instead more complicated.
Nevertheless it is possible to show that, in the absence of excluded  volume
interactions,
the only contributions to $\langle m^2\rangle$
come from the one loop diagrams of fig. \ceq{momentcont}.
In this way  the second topological moment can be exactly computed.
Physically, one is interested to study the topological moments
as a function of the length of the polymers. This requires a slight
modification of the above approach to compute directly the
configurational probability $\tgl$. The details of this calculation
will be presented
elsewhere.
\section{Conclusions}
In the previous Sections the statistical-mechanical problem of an arbitrary
number of topologically entangled polymers has been mapped to a topological
Ginzburg-Landau model, solving in this way a longstanding problem
firstly pointed out in ref. \cite{brsh}. In our model 
the C-S fields acquire a physical significance, since they
propagate the forces acting on the polymers due to the presence of the
topological constraints.
Of course, the topological forces should not distinguish a polymer from
the other, a principle that is apparently
violated by our choice
of coupling constants in eq. \ceq{coefficienti}.
However, explicit calculations show that this symmetry is not spoiled and
actually puts severe constraints on
the number of possible Feynman diagrams to be computed in the perturbative
approach.
For instance, one recognizes from eq. \ceq{effpotol} and fig.
\ceq{momentcont} that the contributions to the one loop effective potential
and to the second topological moment
are completely symmetric in the fields $\Psi_1,\Psi^*_1$ and
$\Psi_2,\Psi^*_2$.
A related problem is the appearance of a spurious parameter in the polymer
free energy \ceq{finalaction}, namely the C-S coupling constant $\kappa$.
Also in this case it is easy to see that the difficulty is only apparent,
since it is possible to remove the dependency on $\kappa$ with
a simple rescaling of the fields
$\Aij$.

In our construction the topological forces
acting on each couple of trajectories $\Gamma_i$ and $\Gamma_j$
are propagated only by the two fields $\Aij$ and $\Bij$.
For that reason, the calculations performed in the case of two polymers can
be extended also to an arbitrary number of polymers
in a straightforward way.
In this way it is possible to conclude from the analysis  of Section III that
the topological interaction
does not change the critical behavior of the system at least in the
one loop approximation.

As already mentioned in the Introduction, the Gauss linking number is a rather
 poor knot invariant.
However, the field theoretical formulation of topologically
linked polymers established here shows a natural way
to
include higher order topological interactions.
This is achieved by adding to the action \ceq{finalaction}
a new contribution of the form:
\beq{
{\cal A}_N^{int}=\sum_{i,j,k=1}^{N-1}\sum_{l,m,n=2\atop i<l;j<m;k<n}^N
\int d^3x\epsilon^{\mu\nu\rho}\left(
\gamma^{ijk}_{lmn}\bA^{(i)}_{(l)\mu} \bA^{(j)}_{(m)\nu}
\bA^{(k)}_{(n)\rho}+\delta^{ijk}_{lmn}\bA^{(i)}_{(l)\mu}
 \bA^{(j)}_{(m)\nu}
\bB^{(k)}_{(n)\rho}+\ldots\right)}{hoints}
Each term appearing in the above sum is topological
and does not vanish apart from special combinations of the indices
$i,j,k,l,m,n$, in which two or three C-S fields coincide.
Clearly, interactions as those given in eq. \ceq{hoints} generate
corrections to the configurational probability $\tgl$ that contain
higher order link invariants. In perturbation theory these corrections
can be computed order by order with techniques similar to those already
exploited in the case of nonabelian C-S field theories
\cite{witten}, \cite{agua}.
To enforce the new topological constraints through Dirac
$\delta-$functions as in eq. \ceq{constraint}, one has to introduce additional
Fourier variables besides the parameters $\Lambda_{ij}$'s.
The possibility of defining the coupling constants
$\gamma^{ijk}_{lmn}$, $\delta^{ijk}_{lmn}$ etc. in terms of these variables
in a way that is suitable to impose relations on  higher order link
invariants
is still an open
problem and deserves more investigations.


\end{document}